\documentclass[pre,twocolumn,showpacs,preprintnumbers,amsmath,amssymb]{revtex4}
\usepackage{simplewick}

\usepackage{graphicx}
\usepackage{dcolumn}
\usepackage{bm}
\newcommand \be{\begin{eqnarray}}
\newcommand \ee{\end{eqnarray}}
\newcommand \ba{\begin{align}}
\newcommand \eea{\end{align}}

\begin{document}
           \csname @twocolumnfalse\endcsname
\title{Theory of water and charged liquid bridges}
\author{K. Morawetz$^{1,2,3}$}
\affiliation{$^1$M\"unster University of Applied Science,
Stegerwaldstrasse 39, 48565 Steinfurt, Germany}
\affiliation{$^2$International Institute of Physics (IIP)
Federal University of Rio Grande do Norte
Av. Odilon Gomes de Lima 1722, 59078-400 Natal, Brazil
}
\affiliation{$^{3}$ Max-Planck-Institute for the Physics of Complex Systems, 01187 Dresden, Germany
}
\begin{abstract}
The phenomena of liquid bridge formation due to an applied electric field is investigated. A new
solution of a charged catenary is presented which allows to determine the static and dynamical stability conditions where charged
liquid bridges are possible. The creeping height, the bridge radius and length as well as the shape of the bridge is calculated showing an asymmetric profile in agreement with observations. The flow profile is calculated from the Navier Stokes equation leading to a mean velocity which combines charge transport with neutral mass flow and which describes recent experiments on water bridges.
\end{abstract}
\maketitle

\section{Introduction}

\subsection{The phenomenon}

\begin{figure}
\includegraphics[width=8cm]{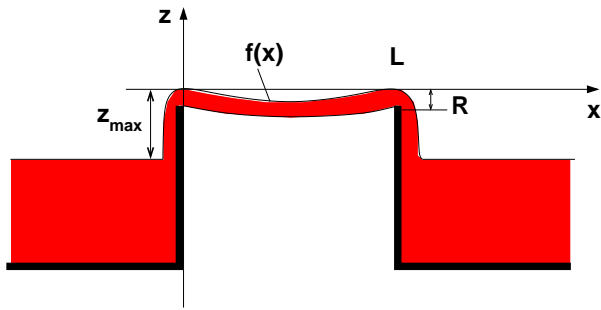}
\caption{\label{scheme} The schematic picture of water bridge between two beakers.}
\end{figure}

The formation of a water bridge between two beakers under high voltage is a phenomenon known since over 100 years \cite{arm1893}. When two
vessels brought in close contact and a high electric field is applied between
the vessels, the water starts creeping up the beakers and forms a bridge which is maintained over a certain distance as schematically illustrated in figure \ref{scheme}. Due to the voltage applied by the vessels the electric field is longitudinally oriented inside the cylindrical bridge. It has
remained attractive to current experimental activities \cite{Wo10,ML10}. On the one side the properties of water
are such complex that a complete microscopic theory of this effect is still lacking. On
the other side the formation of water bridges on nanoscales are of 
interest both for fundamental understanding of electrohydrodynamics  and for 
applications ranging from atomic force microscopy \cite{Sa06} to electrowetting problems \cite{Ju10}. Microscopically the nanoscale wetting is important
to confine chemical reactions \cite{GM06} which reveals an interesting
interplay between field-induced polarization, surface tension, and
condensation \cite{GS03,CZG08}. 

Molecular dynamical simulations have been performed in order to explore the mechanism of
water bridges at the molecular level leading to the formation of
aligned dipolar filaments between the boundaries of nanoscale confinements
\cite{CHVWS10}. A competition was found of
orientation of molecular dipoles and the electric field leading to a threshold
where the rise of a pillar overcomes the surface tension \cite{CZG08}. 
In this respect the
understanding of the microscopic structure is essential to explain such phenomena in
micro-fluidics \cite{Sq05}.
The problem is connected with the dynamics of charged liquids which is
important for capillary jets \cite{Ga97}, current applications in ink printers and electrosprays \cite{Ga10,Hi10}. Consequently
the nonlinear dynamics of breakup of free surfaces and flows has been studied intensively \cite{E93,Egg97}.

Much physical insight can be gained already on the macroscopic scale, where the phenomena of liquid bridging is not restricted to water but can be observed
in other liquids too \cite{Sa10} which shows that it has its origin in
electrohydrodynamics \cite{Me69} rather then in molecular-specific structures.
The traditional treatment is based on the Maxwell pressure tensor where the
electric field effects comes from the ponderomotoric forces and due to
boundary conditions of electrodynamics \cite{LL90}. This is based exclusively on the fact
that bulk-charge states decay on a time scale of the dielectric constant
divided by the conductivity, $\epsilon \epsilon_0/\sigma$, which takes for pure water
$0.14$ms. This decay-time of bulk charges follows from the continuity of
charge density $\dot \rho_c=-\nabla \cdot {\bf j}$
combined with Ohm's law ${\bf j}=\sigma {\bf E}=-\sigma \nabla \phi$ where the
source of the electric field is given by the potential
$\nabla^2\phi=-\rho_c/\epsilon\epsilon_0$. An overview about the different forces
occurring in microelectrode structures are discussed in \cite{Ra98}.

This simple Ohm picture leads to a 
problem in partially charged liquids. Following the Ohm picture one has a constant velocity or 
current density of charged particles caused by the external field and limited
by friction. Contrary,  for incompressible fluids the total mass flux cannot be constant but is dependent on the area
where it is forced to flow through. Both pictures seem to be impossible to reconcile. Here in this 
paper we will present a discussion of this seemingly contradiction leading to a dynamical
stability criterion for the water bridge and a combined flow expression. This
is in line with the idea of \cite{Me69} where the bulk charges have been
assumed to be realized in a surface sheet. While there the migration of
charges to the surface has been considered forming a charged surface sheet, we
adopt here the view point of homogeneously distributed bulk charges which flow
in field direction rather than forming a surface sheet.

In the absence of bulk charges the forces on the water stream are caused by the
pressure due to the polarizability of water described by the high dielectric susceptibility
$\epsilon$. This pressure leads to the catenary form of water bridge 
like a hanging chain \cite{WSSSS09}. While already the simplified model of \cite{Sa10} employing a capacitor picture leads to a critical field strength for the formation of
the water bridge, the catenary model \cite{WSSSS09} has not been reported to
yield such a
critical field. In this paper we will show that even the uncharged catenary
provides indeed a minimal critical field strength for the water bridge
formation in dependence on the length of the bridge. This critical field
strength is modified if charges are present in the bridge which we will
discuss here with the help of a new charged catenary solution. This allows us to explain the asymmetry found in the bridge profile \cite{ML10}.

\subsection{Overview about the paper}

The scenario of water or other dielectric bridges is thought as follows.
Applying an electric
field parallel to two attached vessels the water creeps up the beaker and form a bridge as it is nicely
observed and pictured in \cite{Wo10}. This bridge can be elongated up to a critical field strength and it forms a catenary which becomes asymmetric for higher gravitation to electric field ratios \cite{ML10}. The critical value for stability is sensitively dependent on ion concentrations breaking off already at very low concentrations. The amount of mass flow through the bridge does not follow simple Ohmic transport as we will see in this paper. The schematic picture of the water bridge is given in figure \ref{scheme}.

In this paper we want to advocate the following picture. Imaging a snapshot of the charges flowing through the bridge we cannot decide whether the observed charges are due to static bulk charges or due to the floating motion of Ohmic bulk charges. This flow of charges
within the liquid bridge we can associate with a dynamical bulk charge in the mass
motion which is not covered by the decay of Ohmic bulk charges discussed above. Such a picture is supported by the experimental observation of possible copper ion motion \cite{G09} and by the observation that the water
bridge is highly sensitive to additional external electric fields \cite{Fu07}. Strong fields even create small cone jets \cite{Wo10}. This
dynamical bulk charge will lead us to the necessity to solve the catenary problem including bulk charges. Though
charged membranes have been discussed in the literature \cite{MP09}, a new analytical
solution of the charged catenary is discussed in this paper.

The picture of Ohmic resistors and capacitor as described above
is not sufficient, as one can see from the observation that adding a small amount
of electrolytes to the clean water destroys the water bridge almost immediately. In other words good conducting liquids should not form a water bridge. We will derive an upper bound for charges possibly carried in water in order to remain in stable liquid bridges. 
Though we present all calculations for water parameters summarized in table
\ref{table1}, the theory applies as well to any dielectric
liquid in electric fields.

Four theoretical
questions have to be answered: (i) How is the electric field influencing the
height $z_{\rm max}$ water can creep up? (ii) What is the radius $R(x)$ along the bridge? (iii) What is the form $z=f(x)$ of the water bridge? What are the static constraints on the bridge?
(iv) Which dynamical constraints can be found for possible bridge formation?

\begin{table}
\begin{tabular}{lccc}
density& $\rho$ &= & $10^3$kg/m$^3$\\
dielectric susceptibility& $\epsilon$ &= & 81\\
surface tension& $\sigma_s$ &= & 7.27$\times$$10^{-2}$N/m\\
viscosity& $\eta$ &= & 1.5$\times$$10^{-3}$Ns/m$^2$\\
conductivity of\cr
clean water & $\sigma_0$ &= & 5$\times$10$^{-6}$A/Vm\\
molecular conductivity\cr of NaCl & $\lambda$&=& 12.6$\times$10$^{-3}$Am$^2$/Vmol\\
heat capacity & $c_p$&=&4.187 J/gK
\end{tabular}
\caption{\label{table1} Variables and parameters used within this paper for water.}
\end{table}

We will address all four questions with the help of four parameters composed
of the properties summarized in table \ref{table1} of water. The first one is the
capillary height
\be
a=\sqrt{2 \sigma_s\over \rho g}=3.8 \,{\rm mm}
\label{a}
\ee
with the surface tension $\sigma_s$, the particle density $\rho$ and the gravitational acceleration $g$. The second parameter is the water column height balancing the dielectric
pressure called creeping height in the following
\be
b(E)={\epsilon_0(\epsilon-1) E^2\over  \rho g}=7.22  {\bar E}^2\, {\rm cm}  
\label{b}
\ee
where the dimensionless electric field $\bar E$ is in units of $10^4$V/cm.
The third one is the dimensionless ratio of the force density on the
charges by the field to the gravitational force density
\be
c(\rho_c,E)={\rho_c E\over \rho g}=15.97 {\bar E} {\bar \rho_c}
\label{c}
\ee 
where the charge density ${\bar \rho_c}$ is in units of $ng/l$.
For dynamical consideration the characteristic velocity
\be
u_0={\rho g a^2 \over 32 \eta}\approx 3.02 {\rm m/s}
\label{vel}
\ee
will be useful as the fourth parameter.

The outline of the paper is as follows. In the next section we repeat
shortly the standard treatment of creeping height and bubble radius 
of a liquid but add the pressure by the external electric field on
the dielectric liquid. Then we present the form of the bridge in terms of a
new solution of the catenary equation due to bulk charges in section IV.
In section V we present the flow calculation proposing the picture of moving charged
particles due to the field which drag the neutral particles. This will lead
to a dynamical stability criterion. Then we compare with the experimental data and show the superiority of the present treatment. Summary and conclusion ends up the
discussion in section VI.
 
\section{Answer to question (i): Creeping height and (ii): radius of bridge}
We start to calculate the possible creeping height and use the
pressure tensor for dielectric media \cite{LL90}
\be
\sigma^{ik}=-p\delta_{ik}\!-\!\sigma_s \left ({1\over R_1}\!+\!{1\over R_2}\right )\!+\!\epsilon \epsilon_0 E_i E_k\!-\!\frac 1 2 \tilde
\epsilon \epsilon_0 E^2\delta_{ik} 
\label{press}
\ee
where $p$ is the pressure in the system, $R_1,R_2$ the principal radii of
curvature such that the second term on the right hand side describe the
contribution due to surface tension and the last terms are the parts due
to the forces in the dielectric medium. We assume a density-homogeneous liquid
such that for the dielectric susceptibility $\tilde \epsilon=\epsilon-\rho (d \epsilon/d \rho)_T\approx
\epsilon$. Further we consider first the stationary problem which means that
viscous forces can be neglected in (\ref{press}).

Denoting the components of the normal vector by $e^k$, the
stability condition between water (W) and air (A) is given by
\be
\sigma_{(A)}^{ik}e_{(A)}^k=-\sigma_{(W)}^{ik}e_{(W)}^k=-\sigma_{(A)}^{ik}e_{(W)}^k.
\label{bal}
\ee
Since the principal curvature of the tube is much larger radially than
parallel, we have $R_2\sim \infty$ and denoting the coordinate in the direction of
the height with $z$, the pressure difference
between water and air is $p_W-p_L=\rho g z$. We employ the boundary 
conditions for the normal $E^n$ and tangential $E^t$ components of the electric field
\be
E_{(A)}^n=\epsilon E_{(W)}^n=\epsilon E_n,\qquad E_{(A)}^t=E_{(W)}^t=E_t.
\ee
and the balance (\ref{bal}) with
(\ref{press}) reads
\be
\rho g z +{\sigma_s\over R_1}=\frac 1 2 \epsilon_0(\epsilon-1) (\epsilon
E_n^2+E_t^2).
\label{bal1}
\ee
Please note that due to the migration of charges to the surface one should
consider a surface charge here in principle. We adopt thorough the paper the
simplified picture that the charges remain bulk-like due to the preferred motion
along the field and no surface charges are formed. The influence of such
surface charges is considered as marginal since the curvature of the bridge is
minimal leading to preferential tangential components of electric fields.

\begin{figure}
\includegraphics[width=7cm]{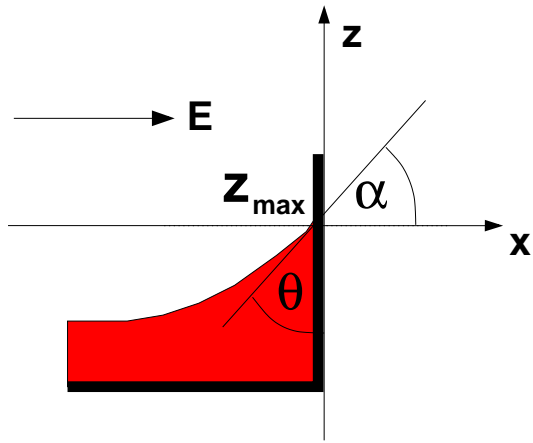}
\caption{\label{scheme1} The schematic picture of water bridge creeping up the
  vessel due to the applied electric field.}
\end{figure}

We assume the electric field in $x$-direction such that
$E_t=-E \cos{\alpha}$, $E_n=E\sin{\alpha}$ where $z'(x)=\tan{\alpha}$ is the
  increase of the surface line of the water as illustrated in figure \ref{scheme1}. Using the parameters
(\ref{a}) and (\ref{b}) we obtain from the stability condition (\ref{bal1})
the differential equation
\be
2 z-a^2{z''\over (1+z'^2)^{3/2}}={\epsilon_0(\epsilon-1)\over \rho g } (\epsilon
E_n^2+E_t^2) \approx b
\label{diffg}
\ee
where we used the approximation of small normal electric fields justified if there are no surface charges.
This shows the modification of the standard treatment of capillary height
by the applied field condensed on the right hand side.  The first integral of (\ref{diffg}) is
\be
{z^2\over a^2}+{1\over \sqrt{1+z'^2}}-{b z\over a^2}=1
\label{int1}
\ee
and we have used the condition that for $x\to\infty$ the surface is 
$z=z'=0$. The explicit
solution of the surface curve $z(x)$ is quite lengthy and not necessary
here. Instead we can give directly the maximally reachable height in
dependence on the electric field. Therefore we use the angle
$\theta=90-\alpha$ of the liquid surface with the wall such that $z'(x)=-\cot{\theta}$ and from (\ref{int1}) we obtain
\be
z=\frac b 2 \!+\!\sqrt{{b^2\over 4}\!+\!a^2 (1\!-\!\sin{\theta})}\le \frac b 2 \!+\!\sqrt{{b^2\over 4}\!+\!a^2}=z_{\rm max} 
\ee
which shows that without electric field the maximal creeping height is  just the capillary length (\ref{a}) as it is well known. The other extreme of very high fields leads to the field-dependent length (\ref{b}) which justifies the name creeping height. This answers the first question concerning creep heights.

The second question, how large the radius of the bridge is, one finds by equating the pressure due to surface tension with the gravitational force density
\be
{\sigma_s\over R}=\rho g z\approx \rho g 2 R 
\ee
such that  the radius of the water bridge is at the beaker
\be
R\approx a/2.
\label{radius}
\ee 
Without using this approximation we could express the curvature again by
differential expressions in $z(x)$ defining a radial profile, as it can be
found in text books \cite{LL90}. The radius of the bridge at the beaker is nearly independent on the applied electric field but only dependent on the surface tension and gravitational force. Along the bridge the radius will change with the applied electric field as we will see later in section IV.C.

\section{Answer to question (iii):  liquid bridge shape}
\subsection{Charged catenary}

Now we turn to the question which form the water bridge will take. Therefore we consider the center of mass line of the bridge being described by $z=f(x)$ with the ends at $f(0)=f(L)=0$. The force densities are multiplied with the area and the length element $ds=\sqrt{1+f'^2} dx$ to  form the free energy. We have the gravitational force density$ \rho g f$ and the volume tension $\rho g b$ as well as the force density by dynamical charges $\rho_c E x$ which contributes. The surface tension is negligible here. The form of the bridge will be then determined by the extreme value of the free energy
\be
&&\int\limits_0^L {\cal F}(x) dx=\rho g \int\limits_0^L (f(x)+b-c x) \sqrt{1+f'^2} dx \to extr.\nonumber\\&&
\label{extr}
\ee
where $c$ is given by (\ref{c}) and $b$ defined in (\ref{b}).

As shown in \cite{M12} and shortly outlines in appendix \ref{solution}
the solution can be represented parametrically as
\be
f(t)&=&{1\over 1\!+\!c^2}\left \{c \,t\!+\!\xi \left [\cosh{\left (\frac t
      \xi\!-\!\frac{ L d}{2\xi}\right )}\!-\!\cosh{\left (L  d\over 2\xi\right
    )}\right ]\right \}\nonumber\\
x(t)&=&t-c f(t),\qquad t\in(0,L).
\label{sol}
\ee
with 
\be
d=2 {\xi \over L}{\rm arcosh}{b\over \xi}
\ee
and $\xi$ to be the solution of the equation
\be
c&=&c_m(\xi,b)\nonumber\\
c_m(\xi,b)\!&=&\!-{2 \xi \over L}\sinh{L\over 2\xi} \!\left ( {b\over \xi} \sinh{L\over 2 \xi}\!-\!\sqrt{{b^2\over \xi^2}-1} \cosh{L\over 2 \xi}\right ).\nonumber\\&&
\label{fa}
\ee

\subsection{Static stability criteria}
Without dynamical bulk charges, $c=0, d=1$, the solution (\ref{sol}) is just the well known catenary \cite{WSSSS09}. The boundary condition (\ref{fa}) reads in this case
\be
{2 b\over L}={2 \xi \over L}\cosh{L\over 2 \xi}\ge \xi_c=1.5088...
\ee
which means that without bulk charges the condition for a stable bridge is
\be
b > \frac 1 2 L \xi_c.
\label{cond1a}
\ee 
Together with (\ref{b}) this condition provides a lower bound for the electric field
in order to enable a bridge of length $L$. This lower bound for an applied field  appears obviously already for the standard catenary  and has been not discussed so far.
 
\begin{figure}
\includegraphics[width=9.5cm]{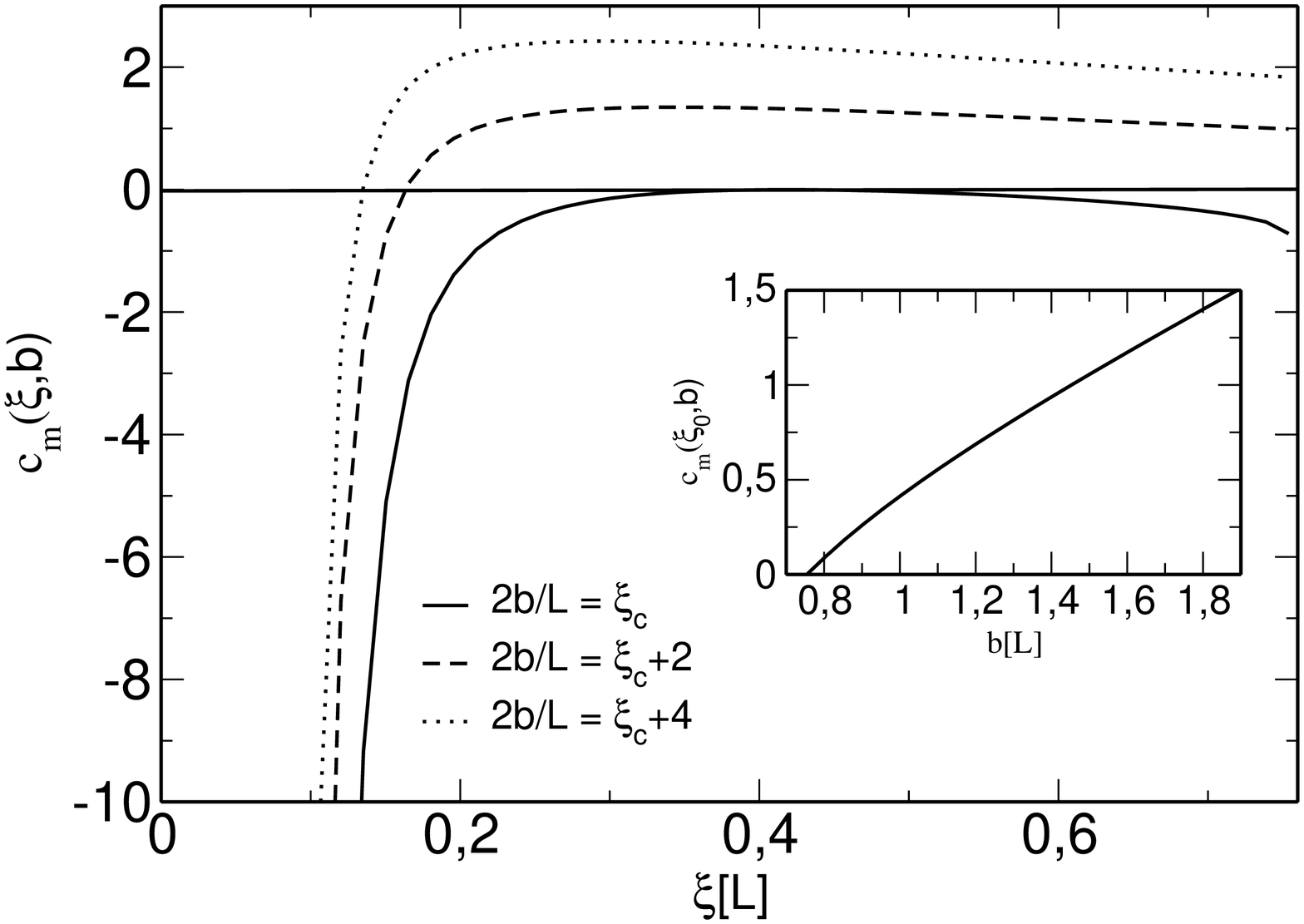}
\caption{\label{figure1} The upper critical bound for the parameter $c$
  according to (\ref{fa}). The inset shows the maximum in dependence on the
  creeping parameter $b$.}
\end{figure}

Lets now return to the more involved case of bulk charges and the new solution of charged catenary (\ref{sol}).
The field-dependent lower bound  condition (\ref{fa}) is plotted in
figure \ref{figure1}. One see that in order to complete (\ref{fa}) the bulk
charge parameter $c$ has to be lower than the maximal value of $c_m$ which reads
\be
c\le c_m(\xi_0,b)
\label{cond1}
\ee
and which is plotted in the inset of figure \ref{figure1}. Remembering the
definition of the bulk charge parameter (\ref{c}) we see that (\ref{cond1})
sets an upper bound for the bulk charge in dependence on the electric field.
The lower bound (\ref{cond1a}) of the electric field 
for the case of no bulk charges is obeyed as well since the curve in the inset
of figure \ref{figure1} starts at $b>L \xi_c/2$ which is the lower bound already present for uncharged catenaries (\ref{cond1a}). 

This completes the third question concerning static stability of the
bridge. We have found a new catenary solution even for bulk charges in the bridge.

\section{Answer to question (iv): Dynamical consideration}
\subsection{Mass flow of the bridge}

We consider now the actual motion of the liquid in the
bridge. Here we want to propose the picture that possible charges in the water
will move according to the applied electric field and will drag water
particles such that a mean mass motion starts. Due to the low
Reynolds numbers (40-100) for water we can consider the motion as laminar and
we can neglect the convection term ${\bf u\nabla} {\bf u}$ in the Navier
Stokes equation \cite{Ch94} which reads then for the stationary case
\be
\eta \nabla^2{\bf u}-\nabla p+\rho_c {\bf E}=0.
\label{Navier}
\ee
The gradient  of the electric pressure (\ref{bal1}) can be given in the direction of the bridge by
\be
-\nabla p={\epsilon_0 (\epsilon-1) E^2\over 2 L}={b\over 2 L}\rho g.
\ee
Here we can adopt the stationary pressure since the viscous pressure is
accounted for by the Navier-Stoke equation.
Assuming that the flow in the bridge has only a transverse component which is
radial dependent, $u(r)$, we can write the Navier Stokes equation (\ref{Navier}) as 
\be
{\eta\over \rho g} {d\over d r} \left (r {d u \over dr}\right )+ r \left
  ({b\over 2 L}+c\right )=0
\ee
with the resulting velocity profile in the direction of the bridge
\be
u(r)-u(R)=2 u_0 \left ({b\over 2 L}+c\right ) \left (1-{r^2\over R^2}\right ) 
\label{ur}
\ee
where $R$ is the radius of the bridge and we have introduced the
characteristic velocity (\ref{vel}). Please note that we keep the undetermined
velocity at the surface of the bridge $u(R)$. We will assume in the follwoing
that it is negligible. The resulting profile (\ref{ur}) has the form of a
Poiseullie flow but with an interplay between forces due to bulk charges and
dielectric pressure in relation to gravity.

The mean current relative to the surface motion is easily calculated  
\be
I=2 \pi \rho \int\limits_0^R dr r [u(r)-u(R)]\equiv \rho v \pi R^2
\ee
providing the mean velocity of the bridge from (\ref{ur}) as
\be
v=u_0 \left ( {b\over 2 L}+c \right ).
\label{v}
\ee
One sees that the ratio of the field-dependent creeping height (\ref{b}) to
the bridge length determines the mean velocity together with possible
dynamical bulk charges described by (\ref{c}). Since we have presently no good control over the surface velocity $u(R)$ we approximate it in the following as zero.

Please note that the bulk charge transport described by (\ref{c}) leads to Ohmic behavior and the neutral particle transport due to dielectric pressure leads to a quadratic field dependence condensed in (\ref{b}). The formula (\ref{v}) combines the effect of charge transport and neutral particle mass transport. It answers the problem raised in the introduction how the two pictures can be brought together, the one of an incompressible fluids where the velocity is dependent of the area and the one of Ohmic transport where the velocity is only dependent on the electric field.

\begin{figure}
\includegraphics[width=9cm]{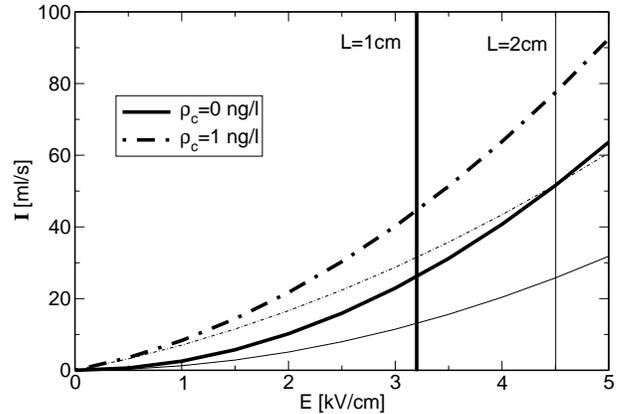}
\caption{\label{current} The mean mass current through the bridge in
  dependence on the electric field and for two different bulk charge
  densities. The thick lines are for a bridge length of $1$cm and the thin
  lines for the corresponding length of $2$cm. The minimal field strength for
  stability (\ref{cond1a}) are indicated by corresponding vertical lines.}
\end{figure}

The resulting total mass
current is given in figure \ref{current}.
The current increases basically with the square of the applied field scaled by
the bridge length. For additional bulk densities the mass flow is higher.

\subsection{Comparison with the experiment}

\begin{figure}
\includegraphics[width=8.5cm]{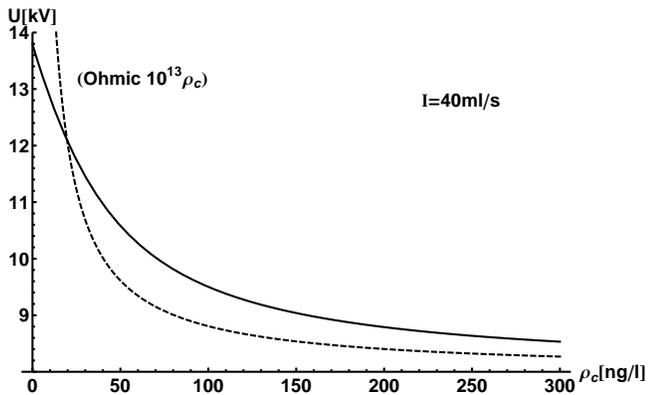}
\caption{\label{massflow} The necessary applied voltage versus bulk charge densities in order to maintain a mass current of $40$ml/s. Following \cite{Wo10} the length of the bridge was $L=1$cm and the diameter $2.5$mm. The result using the flow expression (\ref{v}) of the present paper (solid line) is compared to an Ohmic transport (dashed line). For the latter one the bulk charge has been multiplied with 13 orders of magnitude. }
\end{figure}

To convince the reader about the validity of the velocity formula (\ref{v}) we compare now with the  mass flow and the charge flow measurements.
The experimental values of Figure 4 in \cite{Wo10} are reported to be $40$mg/s for a bridge of $1$cm length, a diameter of $2.5$mm for the stationary regime. 
For this situation we compare in figure \ref{massflow} the results obtained from
(\ref{v}) with a pure Ohmic transport using the lowest-order conductivity expression
\be
\sigma=\lambda {\rho_c \over e N_A} +\sigma_0
\label{bc}
\ee
where for clean water the conductivity is $\sigma_0$, $\lambda$ is the molecular conductivity of the solved charge (electrolyte), and $N_A$  the Avogadro constant, see table \ref{table1}. We see that our formula (\ref{v}) leads to a realistic necessary voltage - which was $12.5$kV in the experiment - even if no bulk charge is presented. In contrast, for the Ohmic transport one has to assume 13 orders of magnitude higher bulk charges to come into the same range. This illustrates the advantage of the here presented model.

\begin{figure}
\includegraphics[width=8.5cm]{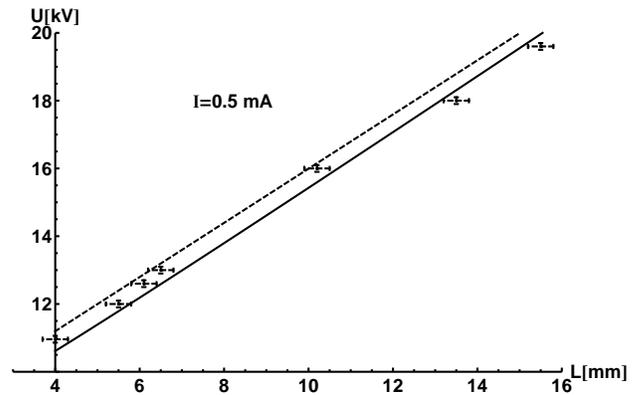}
\caption{\label{chargeflow} The necessary applied voltage versus bridge length in order to maintain a charge current of $0.5$mA. The data are from figure 6 of \cite{Wo10}. The result using the flow expression (\ref{v}) and a bulk charge of $2.3$ng/l (solid line) is compared to an Ohmic transport (dashed line). For the Ohmic transport the bulk charge has been multiplied with a factor of $3\times 10^3$. The same offset of $U_0=8kV$ is used as in the experiments.}
\end{figure}

Considering the charge transport we do not expect such big differences of our model to the pure Ohmic picture since the charged particles matters. To this end we compare the applied voltage versus bridge length with a constant charge current as it was given in figure 6 of \cite{Wo10}. In figure \ref{chargeflow} we compare the result from (\ref{v}) with the pure Ohmic transport. We use a bulk charge of $2.3$ng/l. In order to obtain a comparable Ohmic result we had to multiply the bulk charge with a factor of $3\times10^3$  which illustrates the difference between our model and the Ohmic transport. 

While the difference in charge transport is not very significant provided the fact  that the conductivity of water varies in the order of 3 magnitudes, the mass flow of figure \ref{massflow} has shown that our result here with (\ref{v}) is superior since it considers the drag of neutral particles due to dielectric pressure together with the charge transport.

Having the current at hand one estimates the Joule heating easily as
\be
{\Delta T\over \Delta t}={j E \over \rho c_p}
.
\ee
From figure 5 of \cite{Wo10} one sees that the reported increase of 10K in 30min would translate into field strengths of 0.7kV/cm in our calculation. This is much lower than our result. We would obtain here 2-3 orders of magnitude higher heating rates. Please note that the cooling mechanisms like evaporating and cooling due to water flow is beyond the present consideration. Since these are probably the  major cooling effects in the experiments \cite{Wo11} we cannot compare seriously the theoretical heating rate with the experimentally observed ones.

\subsection{Profile of bridge}
\begin{figure}
\includegraphics[width=8cm]{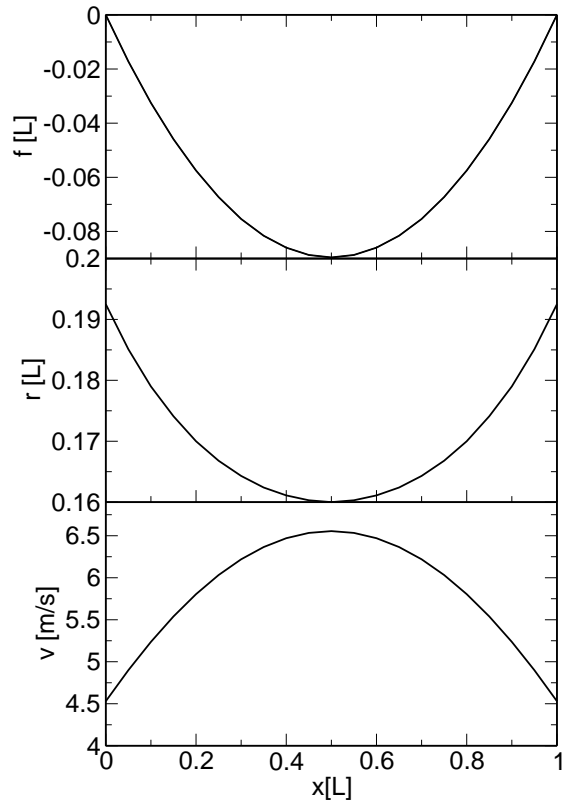}
\includegraphics[width=8cm]{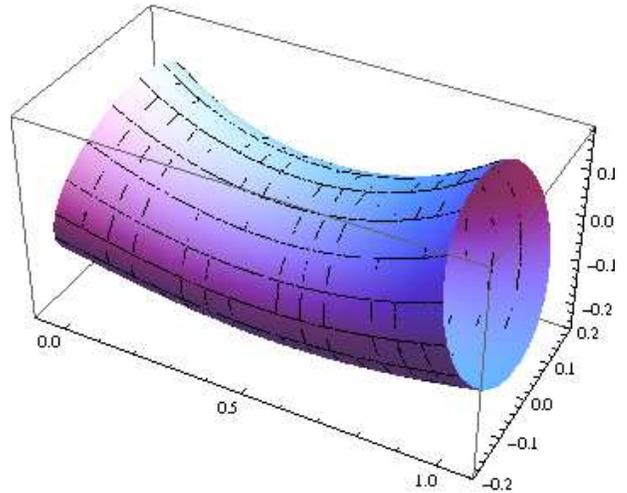}
\caption{\label{figure2} The center of mass coordinate
  (above), the radius (middle) and the velocity (bottom) together with the 3D
  plot  of water bridge (in cm) for no bulk charges $c=0$. The parameter are $b=1.5$cm and according to table \ref{table1}. Please note the different length scales in $x$ and $y,z$ direction.}
\end{figure}

\begin{figure}
\includegraphics[width=8cm]{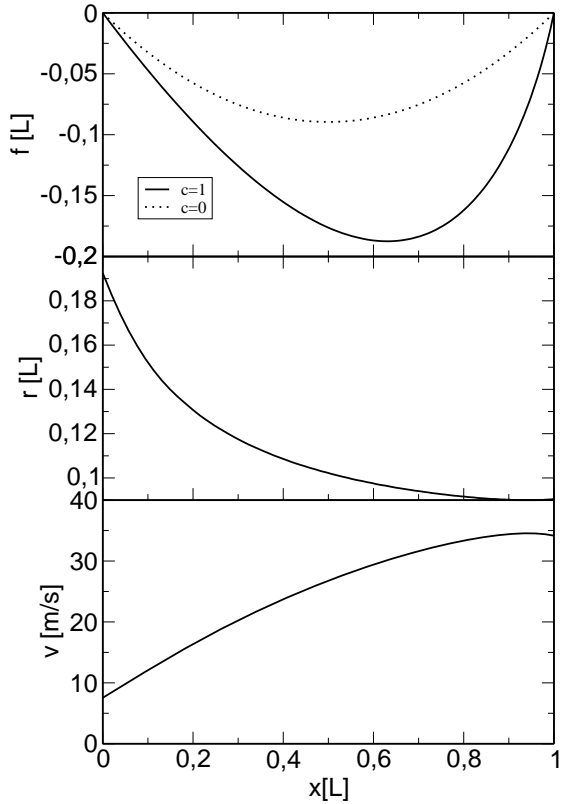}
\includegraphics[width=8cm]{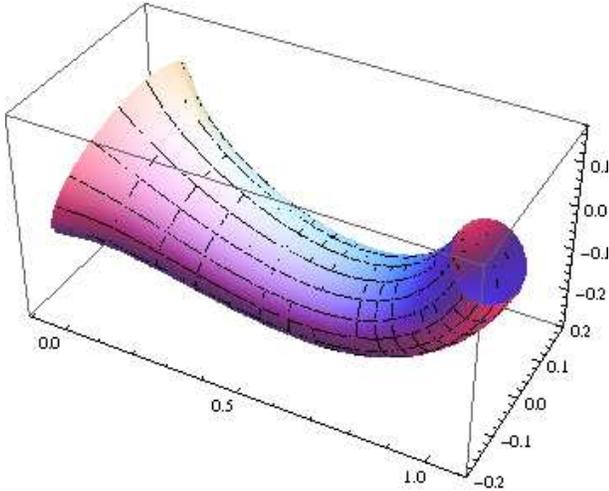}
\caption{\label{figure3} The center of mass coordinate
  (above), the radius (middle) and the velocity (bottom) together with the 3D
  plot  of water bridge (in cm) with bulk charges $c=1$. The parameter are $b=1$cm and according to table \ref{table1}.}
\end{figure}

Let us now calculate the profile of the bridge along the length. We consider
to this end the total mass flow of the bridge and neglect the viscous term compared
to the kinetic energy (which includes part of the  convection term),
${\bf u\nabla} {\bf u}=\frac 1 2 \nabla u^2+{\rm curl} {\bf u}\times {\bf
  u}\approx \frac 1 2 \nabla u^2$. Then one arrives at the Bernoulli equation
\be
\rho {v(x)^2\over 2}\!+\!\rho g f(x)\!+ \!\sigma_s ({1\over R(x)})\!-\!\rho_c E x=\rho {v^2\over 2}\!+\!\sigma_s \frac 1 R.
\label{bern}
\ee 
Here we have neglected the curvature of the bridge compared to the curvature
due to the radius and have compared the position-dependent radius $R(x)$ and
velocity $v(x)$ in the bridge with the situation at the beaker ($x=0$).
The Bernoulli equation (\ref{bern}) can be rewritten in terms of the capillary height (\ref{a}) and the velocity (\ref{v}) as
\be
f(x)-c x={v^2-v^2(x)\over 2 g}+a -{a^2\over 2 R(x)}
\ee
which determines the radius $R(x)$ from the profile of the bridge (\ref{sol})
and the velocity $v(x)$ if we observe the current conservation through an area
\be
R(x)^2 v(x)=R^2 v.
\ee
The results are presented figures \ref{figure2} and \ref{figure3}. We plot the shape of the bridge, the radius and the velocity together with a 3D plot. The case of no bulk charges which leads to the standard catenary can be found in figure \ref{figure2} and figure \ref{figure2} shows the situation for extreme bulk charges almost at the stability edge (\ref{cond1}). We see a deformation of the catenary due to the applied field. This deformation is observed, e.g. if an additional field is brought near the bridge \cite{Fu07,Wo10}. One sees that the radius is becoming smaller at one end of the bridge accompanied with higher velocities as it is known from falling water pipes \cite{HB02}. The bulk charge leads to deformations of this profile which are exaggerated in the plot due to the choice of unequal scales.

Interestingly such asymmetry is experimentally observed \cite{Wo10}, where
after 3 min of operation the asymmetry for the bridge of $0.9$cm length ranges from a diameter of $2.1$mm to $2.6$mm. This is in agreement with the profile calculated in figure \ref{figure3}. Also the measured asymmetry in the left and right catenary angle \cite{ML10} in glycerine can be explained with the present model.
 
\subsection{Dynamical stability}
We turn now to the question of dynamical stability of the flow and consider the motion of water together with the motion of charged particles characterized by the mass $m_i$ and charge $e_i$. This charge current is given by Ohm's law $\sigma E$ and the corresponding mass current can be written
\be
j_i={m_i\over e_i } j=x_i {\rho\over \rho_c} \sigma E
\ee
where we introduced the mass ratio of the number of charged particles (e.g. NaCl) to the water particles
\be
x_i={\#_i m_{{\rm NaCl}}\over \#_w m_{{\rm H}_{2}{\rm 0}}}={\rho_c m_i\over \rho e_i}.
\label{x}
\ee
The mass current of the neutral (water) particles are then
\be
j_n=\rho_n v_n=(\rho-{m_i\over e_i}\rho_c) v_n=(1-x_i) \rho v_n
\ee
such that the total mass current reads
\be
\rho v =j_i+j_n=x_i {\rho\over \rho_c} \sigma E +(1-x_i) \rho v_n.
\ee
The total current (left side) should be larger than the
current only from the charged particles (last term on the right side). However the
velocity of charged particles, $\sigma E/\rho_c$ should be larger than the
velocity of the dragged water molecules $v_n$  and therefore larger than the mean velocity $v$ of the mass motion.  
Together with (\ref{v}) this is expressed by the inequalities
\be
{\sigma E\over \rho_c}>u_0\left ({b\over L}+c\right )>x_i {\sigma E\over \rho_c}
\label{dyn}
\ee
which gives an upper and lower bound on the possible mass motion created by the drag of particles due to the force on charged particles. 

If we now take into account the dependence of the conductivity on the density
of the solved ions in water we can find a condition on possible bulk
charges in water to maintain a stable bridge. To this aim we consider very
small charge densities solved in water which allows to consider the lowest order dependence of the conductivity on the bulk charge concentration (\ref{bc}).

Noting the charge-density dependencies of $x_i$, $b$  and $c$ via (\ref{x}), (\ref{b}) and (\ref{c}) one obtains from (\ref{dyn}) the dynamical restriction on possible bulk charges  
\be
\rho_c&\in& \rho_1-\rho_2\pm\sqrt{(\rho_1-\rho_2)^2+\rho_3^2}\nonumber\\
\rho_c (1-2 \rho_2/\rho_i)&>&\rho_3^2/\rho_i-2 \rho_1
\label{cond2}
\ee
with the auxiliary densities
\be
\rho_1&=&\epsilon_0 (\epsilon-1) {E\over 2 L},\quad \rho_2={16 \eta \lambda \over e N_A a^2}\nonumber\\
\rho_3^2&=&{32 \eta \sigma_0\over a^2},\quad \rho_i={e_i \rho\over m_i}.
\ee

\begin{figure}
 \includegraphics[width=8cm]{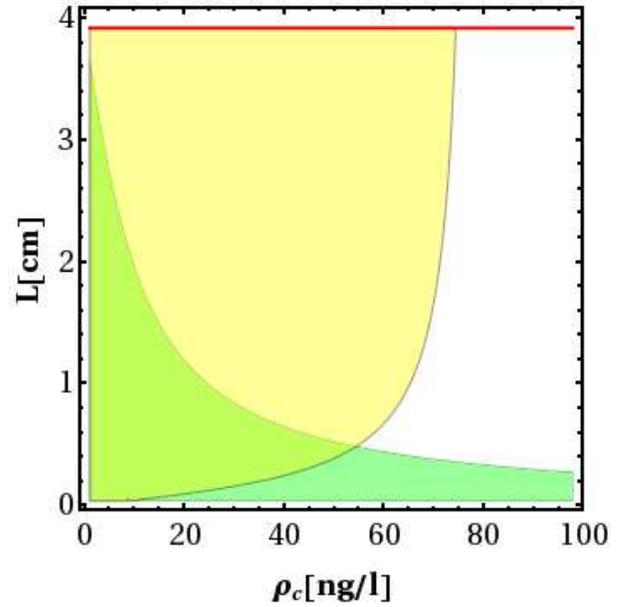}
\caption{\label{figure4} The range of possible water bridges for an electric field of $E=0.64$kV/cm. The upper limit is due to the static stability condition (\ref{cond1}) and the lower cut is due to the dynamical condition (\ref{cond2}). The bulk-charge-free condition is the upper straight line.}
\end{figure}

\begin{figure}
\includegraphics[width=8cm]{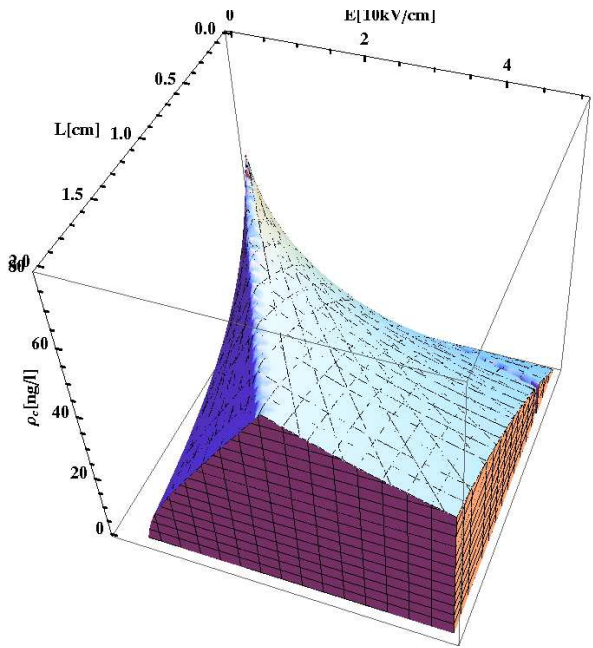}
\caption{\label{figure5} The range of possible water bridges in dependence on the bridge length, the electric field and the electrolyte bulk charges.}
\end{figure}

The results for NaCl in water (table \ref{table1}) are plotted in figures \ref{figure4}-\ref{figure5}. The static stability condition (\ref{cond1a}) gives the upper and charge-density-independent limit in figure \ref{figure5}. The static condition (\ref{cond1}) with bulk charges leads to the border of maximal densities on the right side which agrees with (\ref{cond1a}) at zero densities, of course. The lower minimal length of the bridge at a given field strength and bulk charge is provided by the dynamical condition (\ref{cond2}). For no bulk charge the possible range of lengths of the bridge starts at zero and is limited by the upper length (\ref{cond1a}). If there are charges present, there is a minimal length required to have a stable bridge.   

From the 3D plot in figure \ref{figure5} one can see that for finite charges and for fixed bridge lengths there is a lower and an upper critical field where bridges can only be stable. From the experiments \cite{Wo10} it is seen that the bridge forms jets for fields higher than 15kV/cm and therefore becomes unstable. With a bridge length of $0.5cm$ this translates into a bulk charge of $4$ng/l according to our found boundary conditions. This is in agreement with the value needed to reproduce the flow measurements described in section IV.B.

\section{Summary}
 
The formation of water bridges between two vessels when an electric field is
applied has been investigated macroscopically. Electrohydrodynamics is
sufficient to describe the phenomena in agreement with the experimental
data. The four necessary parameters which are build up from microscopic 
properties of the charged liquid are the capillary height (\ref{a}), the
creeping height (\ref{b}), the dimensionless ratio between field  and
gravitational force density (\ref{c}), and the characteristic velocity (\ref{vel}).

As new contribution to the discussion, an exact solution has been found of a
charged catenary. This leads to a static
stability criterion for possible charges in the liquid dependent on the
applied field strengths and on the length of the bridge. With no bulk charges present the maximal bridge length is determined and no minimal length occurs. This changes if bulk charges are present. Then also a minimal length is required. However, only very small concentrations of bulk charges are possible and the bridge is easily destroyed when bulk charges exceed  50 ng/l.  As a further an asymmetric profile in the diameter along the bridge is obtained which was observed by asymmetric heating.

For the dynamical consideration a
picture is proposed of dragged liquid particles due to the motion of the
charged ones besides the ponderomotoric forces due to the dielectric character
of the liquid. The resulting consideration of dynamical stability restricts the possible
parameter range of bridge formation. The resulting mass flow combines the charge transport and the neutral mass flow dragged by dielectric pressure and is in agreement with the experimental data. 

The presented simple classical 
theory applies for charged liquids as long as the Reynolds number is such low
that laminar flow can be assumed.

\acknowledgments
The discussions with Bernd Kutschan who pointed out this interesting effect to me and the clarifying comments of Jacob Woisetschl\"ager
are gratefully mentioned.
This work was supported by DFG-CNPq project 444BRA-113/57/0-1 and the DAAD-PPP (BMBF) program. The financial support by the Brazilian Ministry of Science 
and Technology is acknowledged.

\appendix
\section{Solution of charged catenary\label{solution}}

Here the drivation of the charged catenary \cite{M12} is shortly sketched.
We solve the variation problem (\ref{extr})
\be
&&\int\limits_0^L {\cal F}(x) dx \to extr.\nonumber\\&&
\ee
with the functional ${\cal F}(x)=\rho g\,   [f(x)+b-c x] \sqrt{1+f'(x)^2}$ and the boundary conditions $f(0)=f(L)=0$.

It is useful to introduce
\be
t(x)= f(x)+b-c x
\ee
such that
\be
{\cal F}(x)=\rho g \,  t(x) \sqrt{1+[t'(x)+c]^2}.
\ee
The corresponding Lagrange equation 
\be
{d\over d x} {\delta {\cal F}\over \delta t'(x)}-{\delta {\cal F}\over \delta t(x)}=0
\ee
possesses a first
integral
\be
t'(x) {\delta {\cal F}\over \delta t'(x)}-{\cal F}={\rm const}=-\xi \sqrt{1+c^2}
\ee
where we introduced the first integration constant $\xi$ in a convenient way. 

The resulting differential equation 
\be
t(\bar x) [c t'(\bar x)+1]=\xi \sqrt{t'(\bar x)^2+(c t'(\bar x)+1)^2}
\ee
with $\bar x =x (1+c^2)$
is solved in an implicit way
\be
t(\bar x)=\xi \cosh\left \{\frac 1 \xi \left [\bar x+c t(\bar x)-c b+\frac L 2 d\right ]\right \}
\ee
with a second integration constants $d$. The profile is therefore given by the implicit equation
\be
f(x)=c x -b+\xi \cosh\left \{\frac 1 \xi \left [x+c f(x)+\frac L 2 d\right ]\right \}.
\label{f1}
\ee

The boundary condition $f(0)=0$ leads to the determination of the integration constant
\be
d=2 {\xi \over L}{\rm arcosh}\left (b\over \xi \right )
\label{dl}
\ee
in terms of the yet unknown $\xi$ constant. The solution (\ref{f1}) can be written with the help of (\ref{dl}) as
\ba
f(x)=c x \!+\!\xi\left \{ \!\cosh\!\!\left [{ x\!+\!cf(x)\over \xi} \!-\!{L d\over 2 \xi} \right] \!-\!\cosh\!\left (\!{L d\over 2 \xi}\! \right )\!\right \}.
\label{f2}
\end{align}
The boundary condition $f(L)=0$ lead to the determination of the remaining constant $\xi$ to be the solution of the equation
\be
c&=&-{2 \xi \over L}\sinh{L\over 2\xi} \left ( {b\over \xi} \sinh{L\over 2 \xi}-\sqrt{{b^2\over \xi^2}-1} \cosh{L\over 2 \xi}\right ).\nonumber\\&&
\label{f3}
\ee
Finally we can rewrite the implicit solution (\ref{f2}) in parametric form. Therefore we choose as parameter $t=x+c f(x)$ which runs obviously through the interval $t\in (0,L)$ and we obtain the solution (\ref{sol})
\be
f(t)&=&{1\over 1\!+\!c^2}\left \{c \,t\!+\!\xi \left [\cosh{\left (\frac t
      \xi\!-\!\frac{ L d}{2\xi}\right )}\!-\!\cosh{\left (L  d\over 2\xi\right
    )}\right ]\right \}\nonumber\\
x(t)&=&t-c f(t),\qquad t\in(0,L).
\ee

\bibliography{bose,kmsr,kmsr1,kmsr2,kmsr3,kmsr4,kmsr5,kmsr6,kmsr7,delay2,spin,spin1,refer,delay3,gdr,chaos,sem3,sem1,sem2,short,cauchy,genn,paradox,deform,shuttling,blase}

\end{document}